\documentstyle[12pt,aaspp4,psfig]{article}

\newcommand{\Ly}{Ly$\alpha$}
\newcommand{\km}{km~s$^{-1}$}

\def\gtorder{\mathrel{\raise.3ex\hbox{$>$}\mkern-14mu
             \lower0.6ex\hbox{$\sim$}}}
\def\ltorder{\mathrel{\raise.3ex\hbox{$<$}\mkern-14mu
             \lower0.6ex\hbox{$\sim$}}}
\def\proptwid{\mathrel{\raise.3ex\hbox{$\propto$}\mkern-14mu

             \lower0.6ex\hbox{$\sim$}}}

\begin{document}

\title{PG 0946+301: the Rosetta Stone of BALQSOs?
}

\author{Nahum Arav\footnote{ To be published 
in the proceednigs of the: 
"Structure and Kinematics of Quasar Broad Line Regions" \\ 
conference, ed.
C. M. Gaskell, W. N. Brandt, M. Dietrich, D. Dultzin-Hacyan, and M.
Eracleous.}
 }
\affil{IGPP, LLNL, L-413, P.O. Box 808, Livermore, CA 94550; \\
I: narav@igpp.llnl.gov }




\begin{abstract}
We describe the motivation and features of a multiwavelength spectroscopic
campaign on broad absorption line (BAL) QSO PG~0946+301.  The main
goal of this project is to determine the ionization equilibrium and
abundances (IEA) in BAL outflows.  Previous studies of IEA in BALQSOs
were based on the assumption that the BALs are not saturated so that
the column densities inferred from the apparent optical depths are
realistic.  This critical assumption is at odds with several recent
observations and with analysis of existing data which indicate that
the absorption troughs are heavily saturated even when they are not
black.  In addition, X-ray observations, which are important for
constraining the ionizing continuum, were not available for those
objects that had UV spectral data.

Quantifying the level of saturation in the BALs necessitates UV
spectroscopy with much higher S/N and broader spectral coverage than
currently exist.  After taking into
account the capabilities of available observatories, our best hope
for a substantial improvement in understanding the IEA in BALQSOs is
to concentrate considerable observational resources on the most
promising object.  Our studies of available HST and ground-based  
spectra show that PG~0946+301 is by far the best candidate for such a
program.  This BALQSO is at least five times brighter, shortward of
1000 \AA\ rest frame, than any other object, and due to its low
redshift it has an especially sparse \Ly\ forest.  At the same time
PG~0946+301 is a typical BALQSO and therefore its IEA should be
representative.  To this effect we are developing a multiwavelength
spectroscopic campaign (UV, FUV, X-ray and optical) on BALQSO PG
0946+301.  We discuss the goals and feasibility
of each observational component: HST, FUSE, ASCA and ground-based.



\end{abstract}




\section{Introduction}

Broad Absorption Line (BAL) QSOs are the main manifestation of AGN
outflows.  BALs are associated with prominent resonance lines such as
C IV $\lambda$1549, Si IV $\lambda$1397, N V $\lambda$1240, and \Ly\
$\lambda$1215.  They appear in about 10\% of all quasars (Foltz et al.
1990) with typical velocity widths of $\sim10,000$ \km\ (Weymann,
Turnshek, \& Christiansen 1985; Turnshek 1988) and terminal velocities
of up to 50,000 \km.  The small percentage of BALQSOs among quasars is
generally interpreted as an orientation effect, and it is probable that
the majority of quasars and other types of AGN harbor intrinsic
outflows (Weymann et al. 1991).  In figure 1 we show the spectrum of
 PG 0946+301 which is a typical high ionization BALQSO.
\vspace{-3cm}
\begin{figure}[h]
 \centerline{
    \psfig{file=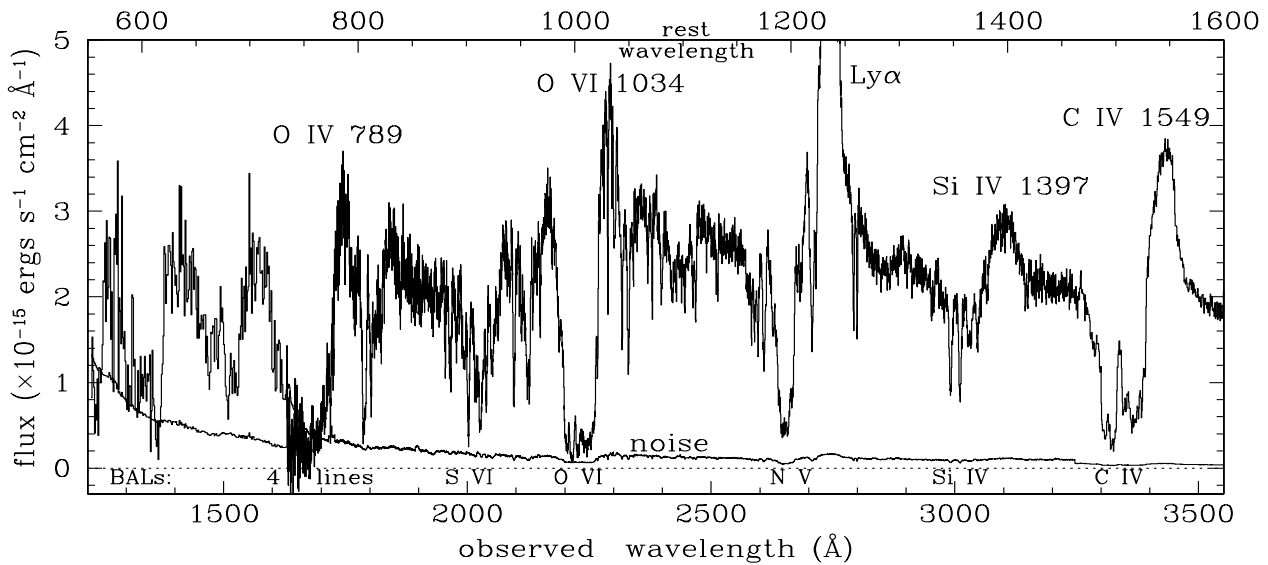,angle=-90,height=12.0cm,width=18.0cm}}
\caption{ HST and ground-based data  for  PG 0946+301 ($z=1.22$), where the
BALs (some of which are  identified at the bottom)
 arise from an outflow with $\Delta v\simeq 10,000$
\km. }\label{0946_spect_full}
\end{figure}

Establishing the physical properties of the flow by determining the
ionization equilibrium and abundances (IEA) of the BAL material is a
fundamental issue in BALQSOs studies.  Furthermore, such
determinations are powerful probes of abundances in the entire AGN
environment.  Inferences about the IEA in the BAL region are derived
by trying to simulate BAL ionic-column-densities ($N_{ion}$) using
photoionization codes.  Several groups (Korista et al. 1996; Turnshek
et al. 1996; Hamann 1996) have used extracted $N_{ion}$ from HST
observations of BALQSO 0226--1024 (Korista et al. 1992) in their IEA
studies while introducing innovative theoretical approaches to the
problem. However, these works used the BAL {\it apparent} optical
depths (defined as $\tau=-ln(I_r)$, where $I_r$ is the residual
intensity seen in the trough) to determine their $N_{ion}$.  The
problem with this approach is that the {\it apparent} optical depths
in the BALs cannot be directly translated to realistic $N_{ion}$
unless the covering factor and level of saturation are known.  Arav
(1997) demonstrated that in BALQSO 0226--1024 the optical depths in
the major BALs are identical within measurement errors.  The
probability of such occurrence by coincidence is less than 1\%, which
strongly suggests that although these BALs are not black, they are all
saturated.  If the BALs are saturated, the inferred $N_{ion}$ are only
lower limits, and thus the conclusions regarding the IEA in this
object (i.e., very high metalicity in the flows; Turnshek et
al. 1996), and by extension in all BALQSOs, are very uncertain.
Evidence for saturation and partial covering factor in the outflows
are also seen in the spectra of Q0449--13 (Barlow 1997) and PG
1254+047 (Hamann 1998), and in spectropolarimetry data (Cohen, M. H.,
et al. 1995; Hines \& Wills 1995).  This, along with the lack of
even a single case where we can say with confidence that saturation is
negligible, makes it likely that saturation affects most (if not all)
observed BALQSOs.

\vspace{.1cm}

Thus it has become evident that an IEA analysis based on apparent
$N_{ion}$ is unreliable and that it is necessary to account for
saturation and partial covering.  Since these effects are
velocity dependent, a detailed study of the optical depth as a
function of velocity [$\tau(v)$] is essential.  Such an analysis
requires much higher S/N data than the apparent $N_{ion}$ approach.
This is currently feasible only for exceptionally UV-bright BALQSOs by
using long integration times.  Observing many objects with a S/N level
similar to that of available data is of limited use for these studies.
In addition to much higher S/N data, a wide UV spectral coverage is
needed to cover lines that arise from the same ion, which yield the
best saturation diagnostics.  A broad UV coverage also yields many lines
from different ions of the same element, which are crucial for
separating effects of ionization and abundances.  These requirements
demand the widest possible spectral coverage and can be achieved by
combining FUSE, HST/STIS and ground based observations. Such
broad-wavelength observations yield more than one BAL for the ions
N~III, O~III, O~IV, Si~IV and S~IV, and BALs from multiple ions for seven
elements (C, N, O, Ne, Mg, Si, S).

\vspace{-.1cm}
\section{Measuring BAL Saturation}
\vspace{-.1cm}

 The best way to measure saturation is by a careful study of different
BALs from the same ion. By doing so we obtain two or more diagnostics
about  precisely the same component of the outflowing gas.  In the case of
PG~0946+301, we have the following  BALs from the same ion
in the combined HST-FUSE spectral coverage: Three N~III BALs 
 associated with transitions at 991 \AA, 685 \AA, and 452 \AA.
Based on their atomic parameters (Verner, Verner, \& Ferland 1996),
the expected optical depth ratio between these line should be 9:20:1,
respectively.  These optical depth ratios give us great dynamical
range in trying to quantify the level of saturation in the lines.
Other ions also have more than one BAL: O~IV 789 \AA, 609 \AA\ and 554
\AA, with $\tau$ ratios of 2:1:5; Si~IV 1397 \AA\ and 458 \AA, with
$\tau_{ratio}=$46:1; S~IV 1070 \AA, 814 \AA, 750 \AA\ and 660 \AA,
with $\tau_{ratio}=1:1.6:11:15$. The case of O~III is somewhat unique
since its lines (835 \AA, 703 \AA\ and 508 \AA) have similar intrinsic
optical depths $\tau_{ratio}=1:1.08:1.05$.  Therefore, these BALs
cannot be used as saturation diagnostics.  Instead they are useful in
fixing the true level of the continuum and for checking the
self-consistency of the $\tau(v)$ solution algorithm.

\vspace{-.1cm}
\section{Analysis of Existing Data}
\vspace{-.1cm}

\begin{figure}[t]
 \centerline{
    \psfig{file=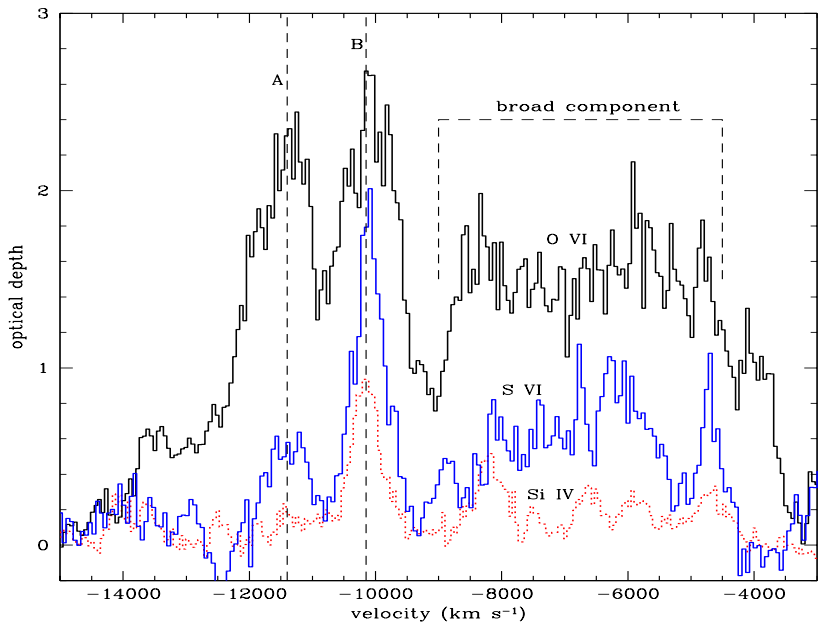,height=9cm,width=12cm}}
\caption{Optical depths as a function of velocity are the important
physical quantities for analyzing the flow. Shown here, on the same
velocity scale, are the $\tau(v)$ for O~VI, S~VI and Si~IV,  where the
main components of the flow are marked. It is clear that these three
$\tau(v)$ are not proportional to each other and therefore using
integrated $N_{ion}$ to study the flow can be very misleading.  For
example by comparing the optical depth ratio between the
high-ionization lines (O~VI and S~VI) and the low-ionization line
Si~IV it is apparent that component A is more highly ionized than
component B.  }\label{3tau}
\end{figure}

Three years ago we began a program to study the IEA in BALQSOs using
HST archival spectra.  We have analyzed the available HST (and
ground-based) spectra of PG~0946+301 (Arav et al.  1998).  Our main
effort was devoted to a detailed study of $\tau(v)$ in the BALs.  For
this analysis we developed a new algorithm to solve for the optical
depth as a function of velocity for doublets and multiplets. $\tau(v)$
for a few lines are shown in figure \ref{3tau}.  We found convincing
evidence for saturation in segments of the troughs, especially in
component B which is seen in all BALs.  This supports our previous
assertion that saturation is common in BALs and therefore cast doubts
on claims for very high metalicity in BAL flows.  We found differing
covering factors for high vs.  low ionization BALs and large
differences in ionization as a function of velocity between different
BALs.  
By comparing the available data of PG~0946+301 to those of all
other BALQSOs in the HST archive, it became obvious that this object
is by far a better candidate for IEA studies than any other known
BALQSO.

%


\vspace{.6cm}

\centerline{\bf Advantages of PG~0946+301}
\vspace{-.2cm}
\begin{itemize}
\itemsep=0pt
\parskip=0pt

\item PG~0946$+$301 is the brightest BALQSO in the UV with 5--10 times
higher flux between 1250--2500\AA\ (observed frame) than the second best
candidate (Q0226--1024).

\item It is a fairly typical BALQSO in terms of luminosity and
optical/UV spectrum and therefore conclusions about its IEA should be
representative of the whole class.

\item The BALs of PG~0946$+$301 are somewhat narrower than those of
other possible candidate, and therefore blend less with each other
across the spectrum.

\item We can obtain very high-quality data for an unprecedented rest frame
spectral region (400 -- 1700~\AA), with very small contamination by \Ly\
forest lines due to the low redshift of the object.

 \end{itemize}

\section{Campaign Description}

Roughly 30 QSO researchers have teamed up in an
international-collaboration dedicated to this multiwavelength
campaign.  Our aim is to obtain: HST UV spectroscopy, FUSE UV
spectroscopy, ASCA X-ray data, high-resolution optical spectroscopy
and optical spectropolarimetry.  Each of these observations will
yield important information by itself, but it is the combined
constraints that will give the most powerful IEA diagnostics.  A 100
ksec ASCA observation has already been approved, a 100 ksec FUSE
proposal has recently been submitted and an HST/STIS proposal of
roughly 40 orbits will be submitted in cycle 8.

\subsection{Observational Components}


{\bf HST}: We expect the highest quality data to come from HST
observations totaling $\sim40$ orbits.  In the far UV (550--800\AA\
rest-frame) a thirty orbit observation with the STIS G140L grating
should give 6-8 times better S/N combined with roughly three times
higher resolution than the available data (shown in Fig. 1).  This
superb data will allow us to extract $\tau (v)$ for 8 important BALs
in that region, which is not possible with the data in hand. Ten
orbits with the G230L grating will give same epoch data with three
times the S/N of the available data between 800--1450 \AA. This
division of observing time should yield data with  S/N$>70$ 
across more than 80\% of  whole HST UV band (1150--3200\AA\ observed-frame). 
 The instrument-related details of the
observations and the data reduction will be handled by Mike Crenshaw
(GSFC) who is a member of the STIS team.

\noindent {\bf FUSE} (Far Ultraviolet Spectroscopic Explorer):
Within the spectral coverage of FUSE (400--530\AA\ in the object's
rest-frame) we expect to find BALs that can serve as unique saturation
diagnostics (Si~IV~$\lambda$458 and N~III~$\lambda$452), which
combined with HST data will enable us to determine saturation levels
up to fifty times the apparent optical depth. BALs from very high
ionization states (Si~XII and S~XIV) will give
information about material in similar ionization states to the ``warm
absorbers'' seen in Seyfert galaxies and quasars.
And BALs from five ions of neon.
The instrument-related details of the observations and the data
reduction will be handled by Mark Giroux (University of Colorado) who
is a part of the FUSE team.

\noindent {\bf ASCA}: X-ray data give us the only direct information about
the shape of the ionizing continuum beyond $\sim$ 30 eV, which is
vital for constraining the ionization equilibrium. Unlike optical/UV
data, the absorption seen in the X-ray data is mainly due to
bound-free opacity, whereas the optical/UV absorption data is due to
bound-bound  opacity (i.e., lines). Therefore,  additional
diagnostics which are much less sensitive to saturation effects
will be obtained by analysis of the X-ray data.  We
have an approved program for a 100 ksec ASCA observation of PG~0946+301
(PI Paul Green).

\noindent {\bf Optical}:
In the optical regime it is important to obtain a very high S/N data
of the C~IV~$\lambda$~1549 BAL (which is the only one observable from the
ground) with a  high spectral resolution.  From these data an accurate
optical depth template can be extracted and compared with all the UV
BALs in order to find ionization differences across the troughs.
It is also beneficial to obtain spectropolarimetry from the ground.
Polarized light gives us information about an indirect photon
trajectory through the BAL region which provides
valuable information about the geometry of the absorbing gas.  


\acknowledgments I thank Kirk Korista, Martijn deKool and Vesa
Junkkarinen for their contributions to this work.  Part of this work
was performed under the auspices of the US Department of Energy by
Lawrence Livermore National Laboratory under Contract W-7405-Eng-48.


%
%

\end{document}